\documentclass[%
 reprint,
superscriptaddress,
 amsmath,amssymb,
 aps,
pra,
longbibliography,
]{revtex4-1}

\usepackage{graphicx}
\usepackage{subfigure}
\usepackage{dcolumn}
\usepackage{bm}

\begin{document}

\preprint{APS/123-QED}

\title{A self-induced mechanism of large-scale helical structures in compressible turbulent flows}

\author{Zheng Yan}
\affiliation{Institute of Applied Physics and Computational Mathematics, Beijing 100094, China}
\author{Jianchun Wang}
\affiliation{Department of Mechanics and Aerospace Engineering, Southern University of Science and Technology, Shenzhen, Guangdong 518055, China}
\author{Lifeng Wang}
\email{wang$\_$lifeng@iapcm.ac.cn}
\affiliation{Institute of Applied Physics and Computational Mathematics, Beijing 100094, China}
\affiliation{Center for Applied Physics and Technology, HEDPS, Peking University, Beijing 100871, China}
\author{Zhu Lei}
\affiliation{Institute of Applied Physics and Computational Mathematics, Beijing 100094, China}
\author{Junfeng Wu}
\affiliation{Institute of Applied Physics and Computational Mathematics, Beijing 100094, China}
\author{Junyi Duan}
\affiliation{LHD, Institute of Mechanics, Chinese Academy of Sciences, Beijing 100190, China}%
\affiliation{School of Engineering Science, University of Chinese Academy of Sciences, Beijing 100049, China}
\author{Fulin Tong}
\affiliation{State Key Laboratory of Aerodynamics, China Aerodynamics Research and Development Center, Mianyang 621000, China}%
\author{Xinliang Li}
\affiliation{LHD, Institute of Mechanics, Chinese Academy of Sciences, Beijing 100190, China}%
\affiliation{School of Engineering Science, University of Chinese Academy of Sciences, Beijing 100049, China}
\author{Changping Yu}
\email{cpyu@imech.ac.cn}
\affiliation{LHD, Institute of Mechanics, Chinese Academy of Sciences, Beijing 100190, China}%
\affiliation{School of Engineering Science, University of Chinese Academy of Sciences, Beijing 100049, China}

\date{\today}

\begin{abstract}
A novel self-sustaining mechanism is proposed for large-scale helical structures in compressible turbulent flows. The existence of two channels of subgrid-scale and viscosity terms for large-scale helicity evolution is confirmed for the first time, through selecting a physical definition of the large-scale helicity in compressible turbulence. Under the influence of the fluid element expansion, it is found that the helicity is generated at small scales via the second-channel viscosity, and the inverse cross-scale helicity transfers at inertial scales through the second-channel helicity flux. Together, they form a self-induced mechanism, which provides a physical insight into the long-period characteristic of large-scale helical structures in the evolution of compressible flow systems.
\end{abstract}

\maketitle


\section{Introduction}
Large-scale helical motions widely exist in natural phenomena and practical applications, such as tornadoes in the atmosphere, Langmuir circulations in the ocean, and Taylor-Görtler vortices in turbomachinery \citep{sanada1993helicity,moffatt1992helicity,yan2020dual}. Nevertheless, these large-scale helical structures have a long-period characteristic, which might originate from an external forcing and intrinsic viscosity. The pseudoscalar of helicity indicates that the intrinsic viscosity might play a self-induced role to decelerate the decay of helicity in some specific circumstances \citep{agoua2021spontaneous}. The existence of large-scale helical structures has an important influence on flow system evolution \citep{teitelbaum2009effect}. This weakens the nonlinearity strength of the Navier-Stokes equations and leads to a smaller viscosity dissipation rate of kinetic energy \citep{bos2013on,linkmann2018effects} and an inverse energy cascade process of kinetic energy \citep{biferale2012inverse,sahoo2017discontinuous}. In engineering flows, large-scale helical structures can strengthen the mixing efficiency in a jet-stirred tubular reactor, decrease aircraft drag, and lead to high heat flux in hypersonic vehicles \citep{moffatt1992helicity,yan2022helicity}. Hence, an induced mechanism should be found to develop appropriate flow control strategies.

Large-scale helical motions can be characterized by high helicity \citep{sanada1993helicity}, and the investigation of large-scale helical structures can be carried out from the perspective of helicity. Helicity is defined as the integral of the scalar product of velocity $\textbf{\textit{u}}(\textbf{\textit{x}},t)$ and vorticity $\bm \omega(\textbf{\textit{x}},t)$ over volume. Consistent with Helmholtz and Kelvin theorems, if we assume that the flows are inviscid and barotropic, the conservation law of helicity is present \citep{moffatt1992helicity}. Helicity measures the twisting, linking and writhing of vortex lines, and it is conservative even in viscous flows \citep{moffatt1969degree,moffatt2014helicity,scheeler2014helicity,scheeler2017complete}. According to the Noether theorem, the conservation law determines the evolution of a three-dimensional turbulent system \citep{pouquet2010the}. It is worth exploring the statistical characteristics of helicity evolution, which is one of the only two inviscid quadratic invariants in three-dimensional flows.

From the perspective of its governing equation, the external sources of helicity generation might be rotating dynamo or buoyancy \citep{agoua2021spontaneous,moffatt1978field,hide1976a,frisch1987large,marino2013emergence}, and the self-induced mechanism might stem from the viscosity term \citep{melander1993polarized}. The role of viscosity in helicity evolution differs from kinetic energy evolution. From the perspective of governing equations, the viscosity plays a dissipation role statistically in the transfer of kinetic energy into internal energy. Nevertheless, viscosity makes a bridge for helicity transferring between different chiralities. Specifically, if viscosity plays a dissipation role for left-chirality helicity, it must play a generating role for right-chirality helicity, and vice versa. Hence, the kinetic energy governing equation must be accompanied by an internal energy governing equation to describe the energy path stringently. However, the helicity governing equation does not need to couple other equations. Besides, viscosity also has other effects \citep{moffatt1992helicity}, such as mediating the reconnection process of vortex lines, causing vorticity to diffuse away from the vortex centerline \citep{kida1988reconnection,irvine2018moreau,yao2020a,yao2021dynamics}. In compressible flows, the role of viscosity is more complex, and compressibility plays an essential role in helicity generation \citep{reshetnyak2012effect,serre2018helicity}. The coupling of the solenoidal and compressible modes leads to a theoretical challenge for exploring the intrinsic mechanism of compressible turbulent flows \citep{john2019solenoidal}. For the existence of large-scale helical structures, there must exist a bridge to transfer helicity from small scales to large scales \citep{kivotides2021helicity,wu2022cascades}.

The unresolved long-period characteristics of the large-scale helical structures in compressible turbulent flows motivate us to comprehensively explore the dynamic evolutions of large-scale helicity. Except for the case-by-case external forcing, we focus on the universal statistical characteristics of helicity evolution at middle and small scales. The rest of the paper is organized as follows. In $\S$ 2, we exhibit the details of DNS of compressible helical turbulent flows. In $\S$ 3, we derive the governing equation of large-scale helicity in compressible turbulent flows, and confirm the existence of dual channel of subgrid-scale (SGS) and viscosity terms of helicity transfer. Next in $\S$ 4, we propose a self-induced mechanism based on the statistical properties of the two channels of SGS and viscosity terms. In $\S$ 5, we summarize the major conclusions and discussions.

\section{Direct numerical simulations of compressible homogeneous and isotropic turbulence flows}

We carry out direct numerical simulations of compressible homogeneous and isotropic turbulent flows in a cubic box with a length of $2\pi$ with a mesh resolution of $2048^3$. The three-dimensional dimensionless Navier-Stokes equations are numerically solved by a hybrid finite difference method, and it combines an eighth-order compact finite difference scheme in smooth regions and a seventh-order weighted essential nonoscillatory scheme in shock regions \citep{wang2013cascade,yan2020cross}. The dimensionless Navier-Stokes equations of ideal gas are
\begin{subequations}
\begin{equation}
\label{eq:mass conservation}
\frac{\partial \rho}{\partial t}+\frac{\partial (\rho u_j)}{\partial x_j}=0,
\end{equation}
\begin{equation}
\label{eq:momentum conservation}
\frac{\partial(\rho u_i)}{\partial t}+\frac{\partial(\rho u_i u_j)}{\partial x_j}=-\frac{\partial p}{\partial x_i}+\frac{1}{Re}\frac{\partial \sigma_{ij}}{\partial x_j}+F_i,
\end{equation}
\begin{equation}
\frac{\partial \mathcal{E}}{\partial t}+\frac{\partial[(\mathcal{E}+p) u_j]}{\partial x_j}=\frac{1}{\alpha}\frac{\partial}{\partial x_j}(\kappa\frac{\partial T}{\partial x_j})+ \frac{1}{Re}\frac{\partial(\sigma_{ij}u_i)}{\partial x_j}- \Lambda+F_j u_j,
\end{equation}
\end{subequations}
where ${\rho}$ is the density, $\textbf{\textit{u}}$ is the velocity, $p$ is the pressure, $T$ is the temperature, $\Lambda$ is a cooling-function for sustaining a steady state statistically \citep{wang2010a}, and $\textbf{\textit{F}}$ is the large-scale force composed of multi-parameters controlling kinetic energy and helicity \citep{yan2020cross}. The multi-parameters include the energy inputting rates of compressive and solenoidal components and helicity inputting rate. Hence, the specific external forcing can be constructed as
\begin{equation}
F_i=\sqrt{\rho}\times(\pi_1 u_i^C + \pi_2 u_i^S + \pi_3 \omega_i),
\end{equation}
where the superscript $C$ and $S$ denote the compressible and solenoidal component of velocity, and $\pi_1$, $\pi_2$ and $\pi_3$ are three indeterminate dimensional parameters. The external forcing is fixed within the lowest two wave number shells to avoid the pollution of external forcing on the statistical results at middle and small scales. Similar to our previous work \citep{yan2020cross}, the injection rates of kinetic energy and helicity are fixed at 0.30 and -0.37, respectively. To obtain the highly compressible turbulent flows, the injection rate of the kinetic energy is divided into the solenoidal and compressible modes equally. More details can refer to our previous work \citep{yan2020cross}.

The equation of state is 
\begin{equation}
p=\rho T/\gamma M^2.
\end{equation}

Besides, the viscous stress $\sigma_{ij}$ and total energy per unit volume $\mathcal{E}$ are defined as 
\begin{equation}
\sigma_{ij}=\mu (\frac{\partial u_i}{\partial x_j}+\frac{\partial u_j}{\partial x_i})-\frac{2}{3}\mu \frac{\partial u_k}{\partial x_k} \delta_{ij}, \mathcal{E}=\frac{p}{\gamma-1}+\frac{1}{2}\rho u_i u_i.
\end{equation}

Some one-point statistical parameters are defined to describe the characteristics of the present compressible helical turbulence \citep{samtaney2001direct,wang2012effect}. The integral length scale $L_f$ is 
\begin{equation}
L_f=\frac{3\pi}{2(u')^2}\int_0^\infty\frac{E(k)}{k}dk.
\end{equation}
The Taylor microscale $\lambda$ and the Kolmogorov length scale $\eta$ are
\begin{equation}
\lambda=\frac{u'}{\langle [(\partial u_1/\partial x_1)^2+(\partial u_2/\partial x_2)^2+(\partial u_3/\partial x_3)^2]/3 \rangle^{1/2}},
\end{equation}
\begin{equation}
\eta=[\langle \mu/Re \rho \rangle^3/\epsilon]^{1/4}.
\end{equation}
Here, $u'$ is the root mean square (rms) of velocity vector $\textbf{\textit{u}}$ which is defined as 
\begin{equation}
u'=\sqrt{\langle u_1^2+u_2^2+u_3^2 \rangle/3}.
\end{equation}

$E(k)$ is the spectrum of kinetic energy per unit mass, $\epsilon$ is the ensemble-averaged viscous dissipation rate of kinetic energy per unit volume which is defined as 
\begin{equation}
\epsilon=\langle \sigma_{ij} S_{ij}/Re \rangle.
\end{equation}
The strain-rate tensor is 
\begin{equation}
S_{ij}=(1/2)(\partial u_i/\partial x_j + \partial u_j/\partial x_i ).
\end{equation}
Hence, the Taylor microscale Reynolds number $Re_\lambda$ and the turbulent Mach number $M_t$ are defined as
\begin{equation}
Re_\lambda=\frac{u'\lambda\langle\rho\rangle}{\langle\mu\rangle}, \qquad M_t=M\frac{\sqrt{3}u'}{\langle\sqrt{T}\rangle}.
\end{equation}

\begin{table*}[htb]
  \caption{Characteristic parameters of steady numerical simulations. Here, $\Delta x$ is the mesh spacing, $\theta'$ is the rms of the velocity divergence and $\omega'$ is the rms of the vorticity magnitude.}
  \label{tab:parameters}
  \begin{ruledtabular}
  \begin{tabular}{cccccccccccccc}
      Resolution  & $Re_\lambda$ & $M_t$ & $\epsilon$ & $\delta$ & $k_{max}\eta$ & $\eta/\Delta x$ & $L_f/\eta$ & $\lambda/\eta$ & $\tau_{\eta}$ & $T_0$ & $\theta'$ & $\omega'$ & $S_3$ \\[3pt]
      \hline
       $2048^3$ & 514 & 0.67 & 0.30 & -0.37 & 3.23 & 1.03 & 477 & 33.96 & 0.032 & 0.94 & 22.22 & 20.47 & -30.76\\
  \end{tabular}
  \end{ruledtabular}
\end{table*}

\begin{figure*}[htb]
	\centering
	\subfigure{
		\begin{minipage}[t]{0.42\linewidth}
			\centering
			\includegraphics[width=\textwidth]{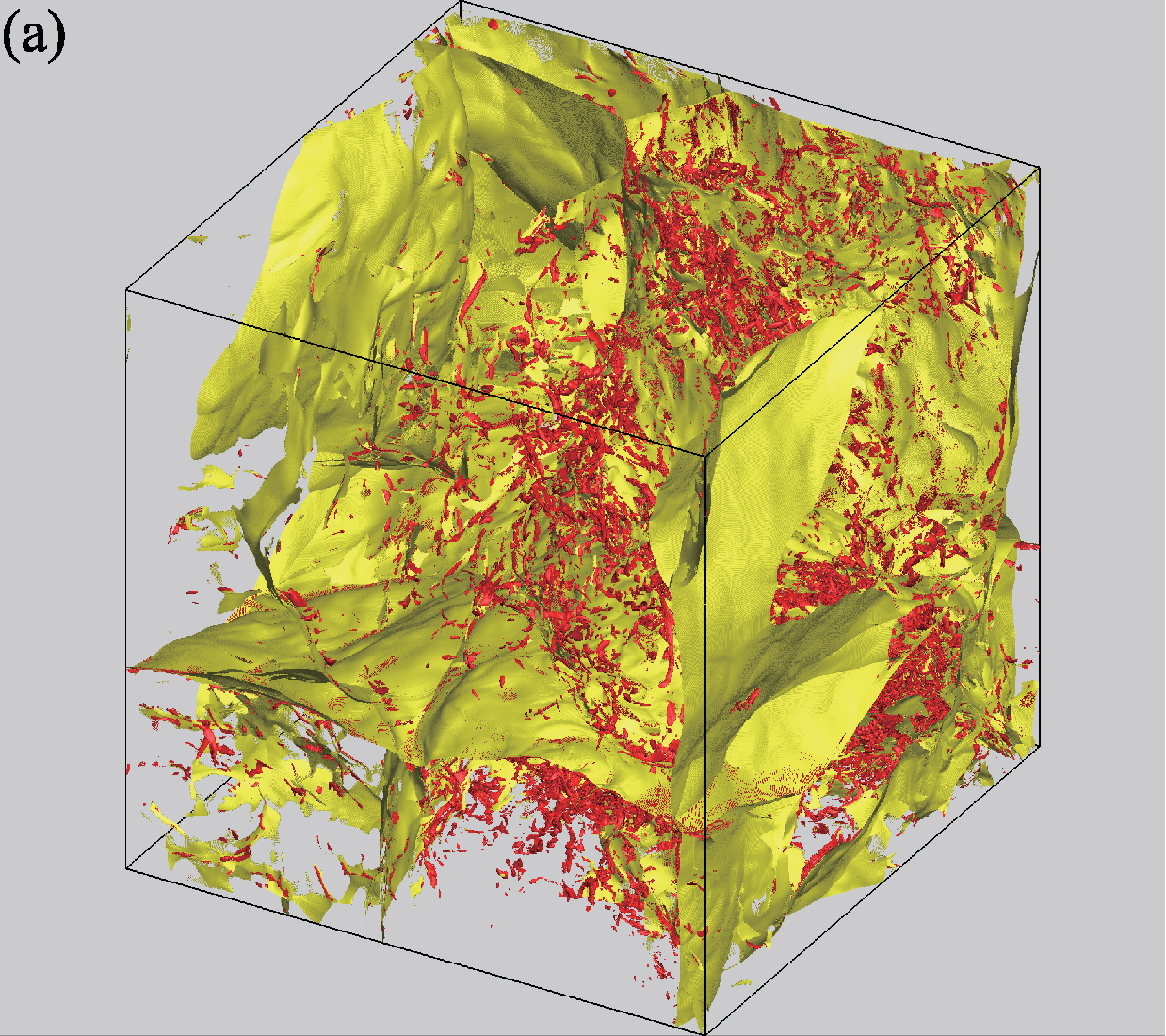}
		\end{minipage}%
	}
	\subfigure{
		\begin{minipage}[t]{0.5\linewidth}
			\centering
			\includegraphics[width=\textwidth]{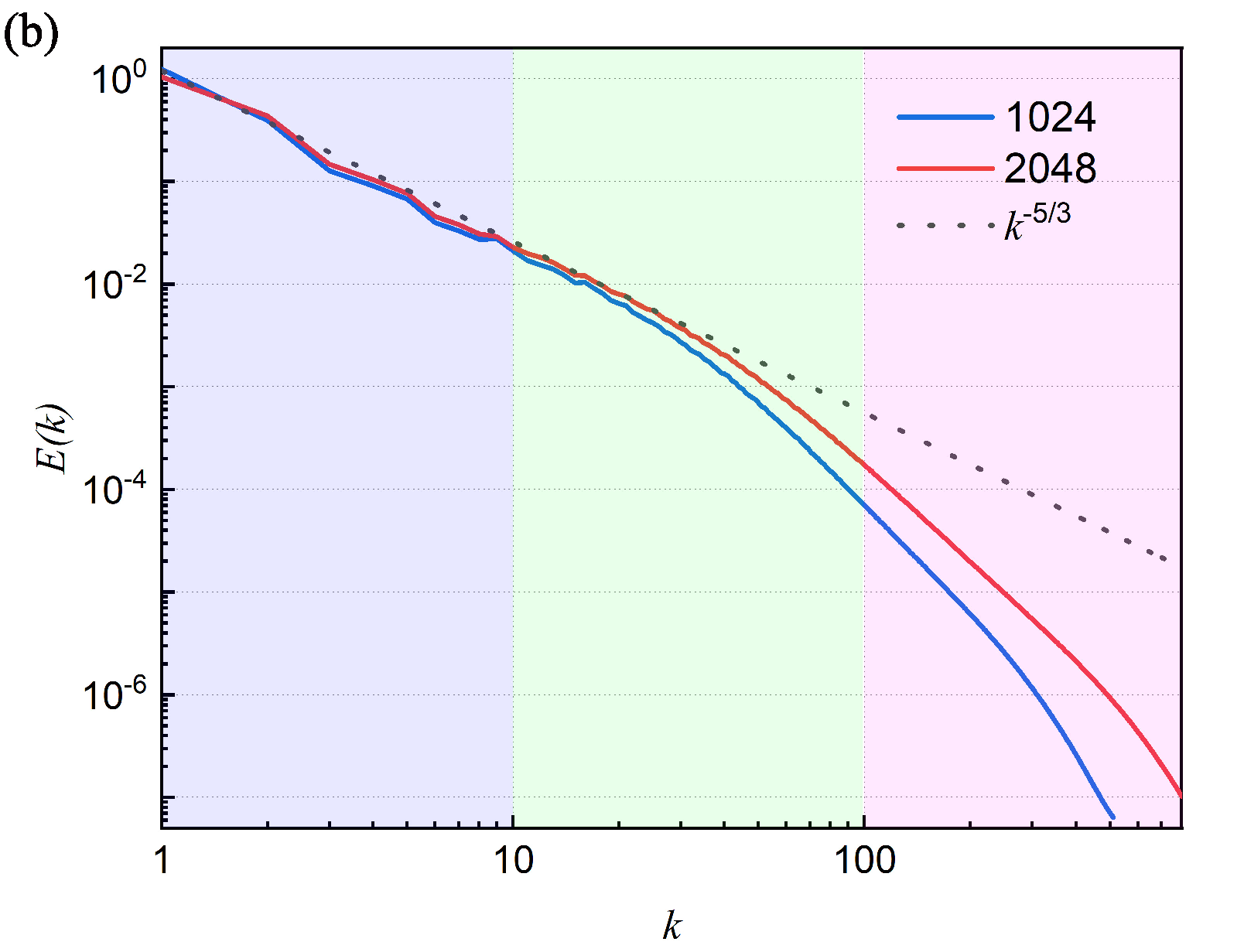}
		\end{minipage}%
	}
	\caption{(a) Isosurface of the velocity divergence with threshold $\theta=-3\theta^\prime$ rendered in yellow and $\theta=0.5\theta^\prime$ rendered in red. (b) Energy spectra of different mesh resolutions, and the energy spectrum of a mesh resolution with $1024^3$ is obtained from our previous work \citep{yan2020cross}.}
	\label{fig:3d_theta_energy_spectrum}
\end{figure*}

The mean viscous dissipation rate of helicity is 
\begin{equation}
\delta=4\left\langle S_{ij}R_{ij}-S_{ii}R_{jj}/3 \right\rangle/Re.
\end{equation}
And $R_{ij}=\left(1/2\right)\left[\partial\left( \omega_i/\rho \right)/\partial x_j+\partial\left( \omega_j/\rho \right)/\partial x_i\right]$. $\tau_{\eta}$ is the Kolmogorov time scale, which is defined as 
\begin{equation}
\tau_{\eta}=\left(\left\langle\mu/\rho\right\rangle/\epsilon\right)^{1/2}.
\end{equation}
$T_0$ is the large eddy turn-over time, which is defined as $T_0=L_f/u'$. The skewness of the velocity gradient $S_3$ is defined as 
\begin{equation}
S_3=\frac{\left[ \left\langle\left(\partial u_1/\partial x_1\right)^3+\left(\partial u_2/\partial x_2\right)^3+\left(\partial u_3/\partial x_3\right)^3\right\rangle \right]/3}{\left[ \left\langle\left(\partial u_1/\partial x_1\right)^2+\left(\partial u_2/\partial x_2\right)^2+\left(\partial u_3/\partial x_3\right)^2\right\rangle /3\right]^{3/2}}.
\end{equation}

The characteristic parameters of the flow fields are listed in table~\ref{tab:parameters}. To obtain the stationary status of fully developed helical turbulent flows, the computational physical time exceeds five times large eddy turnover time. We show the three-dimensional shocklets structures represented by the local velocity divergence of the present mesh resolution in figure~\ref{fig:3d_theta_energy_spectrum}(a). Due to a strong compressibility, the shocklets exhibit large-scale sheet structures, and the expansion structures show a small-scale random distribution. The energy spectra with different mesh resolutions are shown in figure~\ref{fig:3d_theta_energy_spectrum}(b). In contrast to our previous result ~\citep{yan2020cross}, the present energy spectrum with a higher Reynolds number and a higher mesh resolution shows more small-scale structures, which is beneficial to our statistical analysis.

\section{Theoretical derivation}
To study the helicity evolution at different scales simultaneously, we need to derive the governing equation of large-scale helicity in compressible flows. The first undetermined issue is the definition of large-scale helicity because its governing equation should satisfy the conservation law of helicity cascade and inviscid criteria in compressible flows \citep{zhao2018inviscid}. This means that no other source terms exist other than the large-scale pressure, SGS and viscosity terms \citep{yan2019effect}. The definition based on the Favre filter is appropriate, and it is defined as $\widetilde{H} =\widetilde{\textbf{\textit{u}}}\cdot\widetilde{\bm \omega}$. In contrast, the definition based on the direct filter ( $\overline{H}=\overline{\textbf{\textit{u}}}\cdot\overline{\bm \omega}$) might not be appropriate because of existing source terms, and the specific derivation is provided in Appendix~\ref{app:direct definition of large-scale helicity}. Here, $\widetilde{\textbf{\textit{u}}}(\textbf{x})=\overline{\rho \textbf{\textit{u}}(\textbf{\textit{x}})}/\overline{\rho}$ is the Favre-filtered velocity, and $\widetilde{\bm \omega}=\nabla\times\widetilde{\textbf{\textit{u}}}$. $\overline{\cdot}$ denotes a filter operation, which is defined as $ \overline{\textbf{f}}(\textbf{x})=\int{{\rm d}^3\textbf{\textit{r}} G_l(\textbf{\textit{r}}) \textbf{f} (\textbf{\textit{x}}+\textbf{\textit{r}})}$ with kernel $G_l(\textbf{\textit{r}})$.

Making a filter operation on equation~\ref{eq:momentum conservation} and dividing by $\overline{\rho}$, we can obtain the following large-scale velocity equation as 
\begin{equation}
\label{eq:favre filtered velocity equation}
\frac{\partial \widetilde{u}_i}{\partial t}+\widetilde{u}_j\frac{\partial \widetilde{u}_i}{\partial x_j}=-\frac{1}{\overline{\rho}}\frac{\partial \overline{\rho}\widetilde{\tau}_{ij}}{\partial x_j}-\frac{1}{\overline{\rho}}\frac{\partial \overline{p}}{\partial x_i}+\frac{1}{Re}\frac{1}{\overline{\rho}}\frac{\partial \overline{\sigma}_{ij}}{\partial x_j}+\frac{\overline{F}_i}{\overline{\rho}},
\end{equation}
where Favre-filtered SGS stress is $\widetilde{\tau}_{ij}=\widetilde{u_i u_j}-\widetilde{u}_i\widetilde{u}_j$. Making a curl operation on equation~\ref{eq:favre filtered velocity equation}, we can obtain the following Favre-filtered vorticity governing equation as 
\begin{equation}
\label{eq:favre filtered vorticity equation}
\begin{split}
\frac{\partial \widetilde{\omega}_i}{\partial t}+\widetilde{u}_j\frac{\partial \widetilde{\omega}_i}{\partial x_j} & +\widetilde{\omega}_i \widetilde{\theta}= \widetilde{\omega}_j\frac{\partial \widetilde{u}_i}{\partial x_j}-\epsilon_{ijk}\frac{\partial}{\partial x_j}\left( \frac{1}{\overline{\rho}}\frac{\partial \widetilde{\tau}_{kl}}{\partial x_l} \right) \\ & -\epsilon_{ijk}\frac{\partial}{\partial x_j}\left( \frac{1}{\overline{\rho}}\frac{\partial \overline{p}}{\partial x_k} \right)  +\frac{1}{Re}\epsilon_{ijk}\frac{\partial}{\partial x_j}\left( \frac{1}{\overline{\rho}}\frac{\partial \overline{\sigma}_{kl}}{\partial x_l} \right) \\ & +\epsilon_{ijk}\frac{\partial}{\partial x_j}\left( \frac{\overline{F}_k}{\overline{\rho}}\right).
\end{split}
\end{equation}
Hence, combining equation~\ref{eq:favre filtered velocity equation} and ~\ref{eq:favre filtered vorticity equation}, and we can get the following governing equations of large-scale helicity $\widetilde{H}$ with a Favre filter,
\begin{equation}
\label{eq:favre filtered helicity equation1}
\frac{\partial \widetilde{H}}{\partial t}+\frac{\partial \widetilde{J}_j}{\partial x_j}=-\Pi^{H1}_{l}-\Pi^{H2}_{l}+\Phi_{l}^1+\Phi_{l}^2+D_{l}^1+D_{l}^2+\mathcal{F}_{l}^1+\mathcal{F}_{l}^2,
\end{equation}
where $\widetilde{J}_j$ is the spatial transport term, which is defined as
\begin{subequations}
\begin{equation}
\begin{split}
\widetilde{J}_j = & \widetilde{u}_j \widetilde{H}-\left( 1/2 \right)\widetilde{u}_i \widetilde{u}_i \widetilde{\omega}_j + \widetilde{\omega}_i \widetilde{\tau}_{ij} + \epsilon_{ikl}\frac{\partial \overline{\rho} \widetilde{\tau}_{lj}}{\partial x_k}\frac{\widetilde{u}_i}{\overline{\rho}} \\ & +\overline{\rho} \widetilde{\tau}_{ij}\epsilon_{ikl} \widetilde{u}_k \frac{\partial}{\partial x_l}\left(\frac{1}{\overline{\rho}}\right) +\overline{p}\frac{\widetilde{\omega}_j}{\overline{\rho}} +\overline{p}\epsilon_{jkl}\frac{\partial \overline{\rho}}{\partial x_k}\frac{\widetilde{u}_l}{\overline{\rho}^2} \\ & -\overline{\sigma}_{ij}\frac{\widetilde{\omega}_i}{\overline{\rho}}  +\overline{\sigma}_{ij}\epsilon_{ikl}\frac{\widetilde{u}_k}{\overline{\rho}^2}\frac{\partial \overline{\rho}}{\partial x_l}-\frac{\widetilde{u}_i}{\overline{\rho}}\epsilon_{ikl}\frac{\partial\overline{\sigma}_{lj}}{\partial x_k}.
\end{split}
\end{equation}
The other terms are defined as
\begin{equation}
\label{eq:first channel}
\Pi^{H1}_{l}=-\overline{\rho} \widetilde{\tau}_{ij}\frac{\partial}{\partial x_j}\left(\frac{\widetilde{\omega}_i}{\overline{\rho}}\right), \quad \Pi^{H2}_{l}=\Pi^{H21}_{l}+\Pi^{H22}_{l},
\end{equation}
\begin{equation}
\Pi^{H21}_{l}=-\epsilon_{ikl}\frac{\partial \overline{\rho} \widetilde{\tau}_{lj}}{\partial x_k}\frac{\partial \left(\widetilde{u}_i/\overline{\rho}\right)}{\partial x_j},
\end{equation}
\begin{equation}
\Pi^{H22}_{l}=-\overline{\rho} \widetilde{\tau}_{ij}\frac{\partial}{\partial x_j}\left(\epsilon_{ikl} \widetilde{u}_k \frac{\partial \left(1/\overline{\rho}\right)}{\partial x_l}\right),
\end{equation}
\begin{equation}
\label{eq:filtered pressure terms}
\Phi^1_l=\overline{p}\frac{\partial}{\partial x_j}\left( \frac{\widetilde{\omega}_j}{\overline{\rho}} \right), \quad \Phi^2_l=\overline{p}\frac{\partial}{\partial x_j}\left( \epsilon_{jkl}\frac{\partial \overline{\rho}}{\partial x_k}\frac{\widetilde{u}_l}{\overline{\rho}^2} \right),
\end{equation}
\begin{equation}
\label{eq:filtered viscous terms1}
D^1_l= -\overline{\sigma}_{ij}\frac{\partial}{\partial x_j}\left(\frac{\widetilde{\omega}_i}{\overline{\rho}}\right), \quad D^{2}_{l}=D^{21}_{l}+D^{22}_{l},
\end{equation}
\begin{equation}
\label{eq:filtered viscous terms}
D^{21}_l=-\epsilon_{ikl}\frac{\partial\overline{\sigma}_{lj}}{\partial x_k}\frac{\partial}{\partial x_j}\left(\frac{\widetilde{u}_i}{\overline{\rho}}\right), \quad D^{22}_l=\overline{\sigma}_{ij}\frac{\partial}{\partial x_j}\left( \epsilon_{ikl}\frac{\widetilde{u}_k}{\overline{\rho}^2}\frac{\partial \overline{\rho}}{\partial x_l} \right).
\end{equation}
\begin{equation}
\mathcal{F}_{l}^1=\widetilde{\omega}_i\frac{\overline{F}_i}{\overline{\rho}},\qquad \mathcal{F}_{l}^2=\widetilde{u}_i\epsilon_{ijk}\frac{\partial}{\partial x_j}\left( \frac{\overline{F}_k}{\overline{\rho}}\right).
\end{equation}
\end{subequations}

The above governing equation of the large-scale helicity describes the evolution of helical structures in compressible turbulent flows via different physical processes, such as the spatial transportation process ($\widetilde{J}_j$), the SGS transfer ($\Pi^{H1}_{l}$ and $\Pi^{H2}_{l}$), pressure process ($\Phi^{1}_{l}$ and $\Phi^{2}_{l}$), viscosity process ($D^{1}_{l}$ and $D^{2}_{l}$) and external forcing ($\mathcal{F}^{1}_{l}$ and $\mathcal{F}^{2}_{l}$). For the different physical processes, all of them consist of two expressions. To distinguish these two expressions, we name the expression from the momentum equation as the first channel, and name the expression from the vorticity equation as the second channel. These two channels describe the large-scale helicity evolutions from different perspectives, and we investigate their statistical characteristics subsequently.

Here, for the SGS terms, if we assume that the filtered density is a constant, the first and second  SGS channels are recovered to the first and second channels in incompressible flows\citep{yan2020dual}. Notably, the second term of the second-channel helicity flux depends on the density gradient, and it is only present in variable-density flows. In addition, their tensor geometries are more complex than that in incompressible form, and the specific illustration is provided in Appendix~\ref{app:first-term tensor geometry}. For the above two pressure terms, we take the difference and obtain the following relation
\begin{equation}
\begin{split}
\Theta & \equiv \overline{p}\frac{\partial}{\partial x_j}\left( \epsilon_{jkl} \frac{1}{\overline{\rho}} \frac{\partial \widetilde{u}_l}{\partial x_k} + \epsilon_{jkl}\frac{\partial }{\partial x_k}\left( \frac{1}{\overline{\rho}} \right)\widetilde{u}_l \right) \\ & \equiv \overline{p}\frac{\partial}{\partial x_j}\left( \epsilon_{jkl}\frac{\partial }{\partial x_k}\left( \frac{1}{\overline{\rho}} \widetilde{u}_l \right)  \right) \overset{\nabla \cdot \nabla \times \textit{\textbf{f}} \equiv 0 }{ \equiv} 0,
\end{split}
\end{equation}
where $\textit{\textbf{f}}$ is a vector. Hence, the first and second channels of the pressure terms are the same, which means that $\Phi^1_l=\Phi^2_l$. Regarding the statistical characteristics of the pressure terms of helicity transfer in compressible turbulent flows, we performed a detailed analysis in our previous paper \citep{yan2019effect}. The viscosity terms under a constant-density hypothesis can be simplified as
\begin{equation}
D^1_l=-2\nu\overline{S}_{ij}\overline{R}_{ij}, \qquad D^2_l=-2\nu\overline{R}_{ij}\overline{S}_{ij}.
\end{equation}
Under the hypothesis of constant density, the Favre filter can be reduced to other direct filters, and we replace the symbol $\widetilde{\cdot}$ with the symbol $\overline{\cdot}$ in the above derivation. Here, $\overline{S}_{ij}=\left( \partial \overline{u}_i/ \partial x_j  + \partial \overline{u}_j/ \partial x_i\right)/2$ is the symmetric component of the velocity gradient tensor, and $\overline{R}_{ij}=\left( \partial \overline{\omega}_i/ \partial x_j  + \partial \overline{\omega}_j/ \partial x_i\right)/2$ is the symmetric component of the vorticity gradient tensor. We can conclude that $D^1_l=D^1_2$ in incompressible turbulent flows. Hence, the dual-channel characteristics of the viscosity term do not exist in incompressible turbulent flows, and it is consistent with our previous work \citep{yan2020dual}. However, the equality relationship of the dual channels of the viscosity term in compressible turbulent flows does not exist. Therefore, we conclude that the dual-channel characteristics of viscosity terms only exist in compressible turbulent flows.

$\mathcal{F}_{l}^1$ and $\mathcal{F}_{l}^2$ are the external forcing terms. In our numerical simulations, we construct an external forcing consisting of velocity and vorticity within the first two wave number shells to sustain a stationary highly helical turbulent flow. Because we focus on the statistical characteristics of helicity transfer at middle and small scales, the statistical analysis of the external forcing will not be carried out. However, in practical applications, the unknown external forcing might have the dual-channel characteristic.

\section{A self-induced mechanism of large-scale helical structures}
\begin{figure}[htb]
	\centering
	\subfigure{
		\begin{minipage}[t]{0.9\linewidth}
			\centering
			\includegraphics[width=\textwidth]{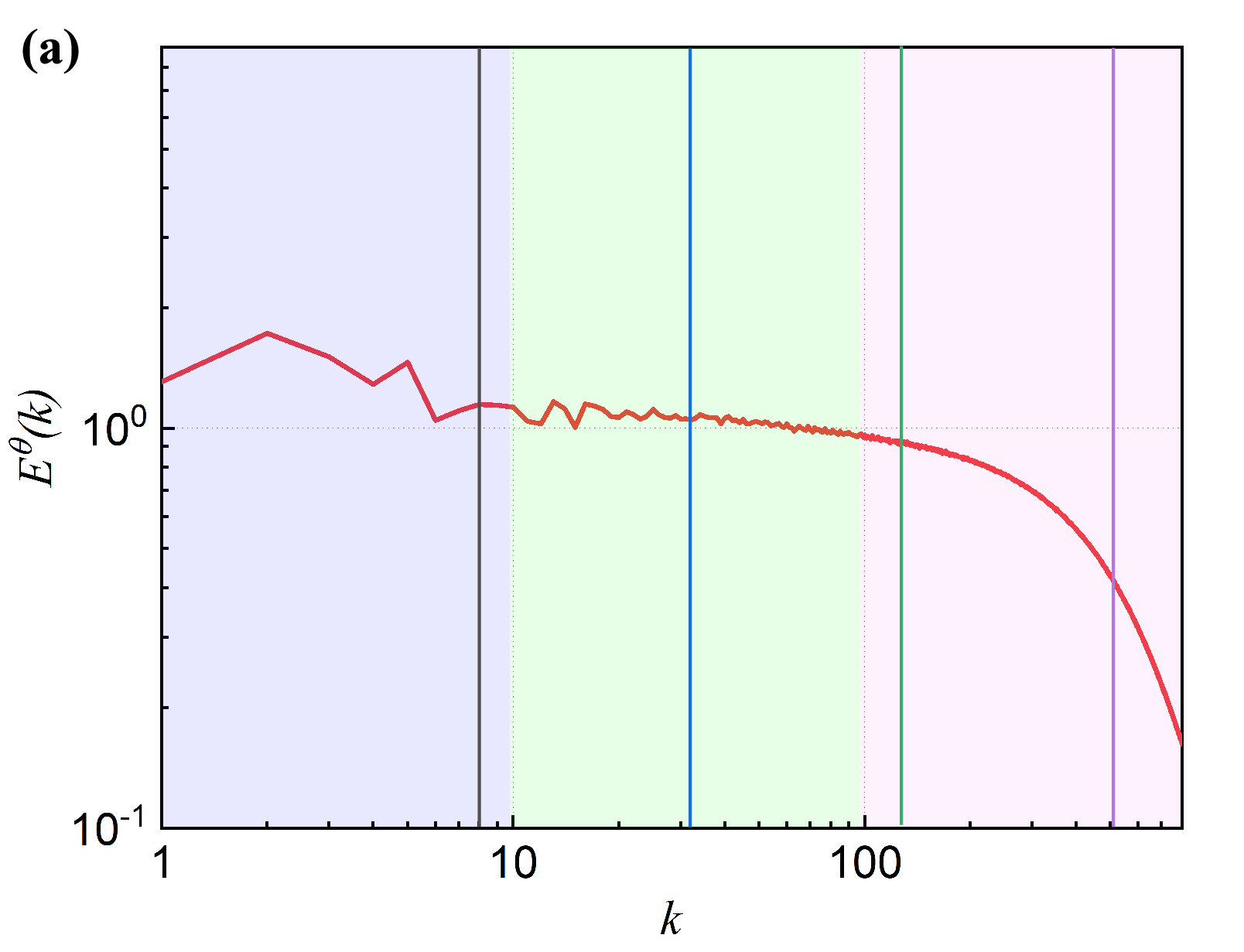}
		\end{minipage}%
	}
	
	\subfigure{
		\begin{minipage}[t]{0.9\linewidth}
			\centering
			\includegraphics[width=\textwidth]{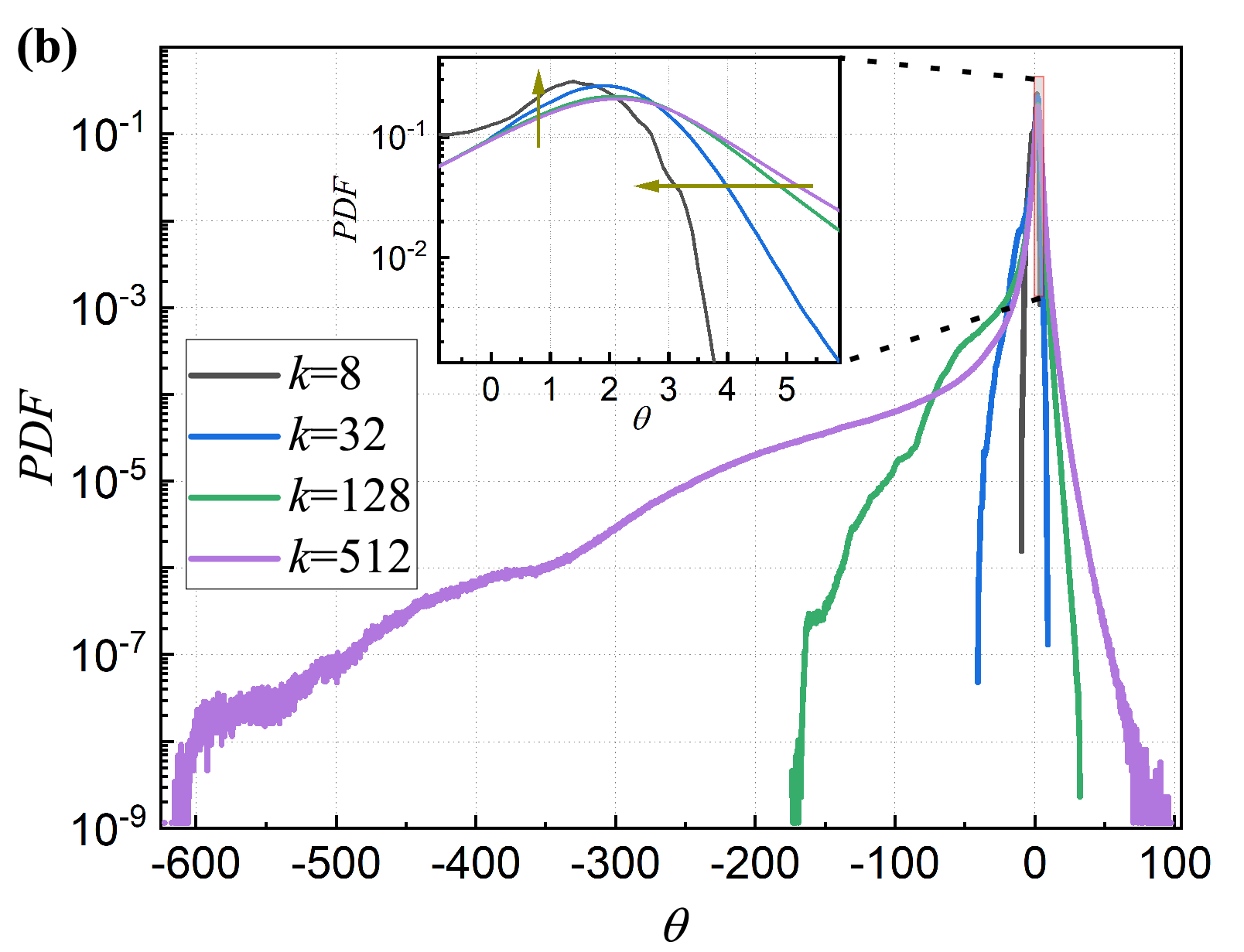}
		\end{minipage}%
	}
	\caption{(a) Power spectrum of the velocity divergence $\theta$. (b) PDF of the velocity divergence at different length scales, and the line color corresponds to the length scales marked up in (a). The inset shows the local enlarged PDF of the velocity divergence to highlight the distribution of the dilatation at different length scales. }
	\label{fig:dilatation}
\end{figure}

The conditional averaging method is used to evaluate the local compression or expansion effects on helicity transfer, and the sufficient sample points in our numerical simulations verify its accuracy \citep{block2006prospects,wang2012scaling,wang2013cascade}. Before evaluating the compression or expansion effects, we need to investigate the scale distributions of the velocity divergence. The power spectrum of the velocity divergence $\theta$ is defined as $\int_0^{\infty}E^{\theta}(k) dk=\left\langle \theta^2 \right\rangle$. In figure~\ref{fig:dilatation}(a), we show the power spectrum of the velocity divergence. The power spectrum density indicates the slight discrepancy of the fluctuations of the velocity divergence at different length scales, especially at large and middle scales. However, both the compression and expansion motions are involved in the power spectrum of the velocity divergence. In order to access the compression and expansion motions separately, we exhibit the probability distribution functions (PDF) of the velocity divergence at corresponding length scales in figure~\ref{fig:dilatation}(b) marked up in figure~\ref{fig:dilatation}(a). The PDF of the velocity divergence indicates that the compression motion is dominant at all scales. With the decrease of the length scale, the dominant degree increases, which is associated with the shocklet structures. For the PDF of the expansion motion shown in the inset of figure~\ref{fig:dilatation}(b), weak expansion motions are more concentrated at middle and large length scales. Hence, the degree of compression and expansion motions is high at middle and small scales in our numerical simulations. This is beneficial for accessing the compression and expansion effects on the SGS transfer at middle length scales and on the viscosity dissipation at small scales.

\begin{figure}[htb]
	\centering
	\subfigure{
		\begin{minipage}[t]{0.9\linewidth}
			\centering
			\includegraphics[width=\textwidth]{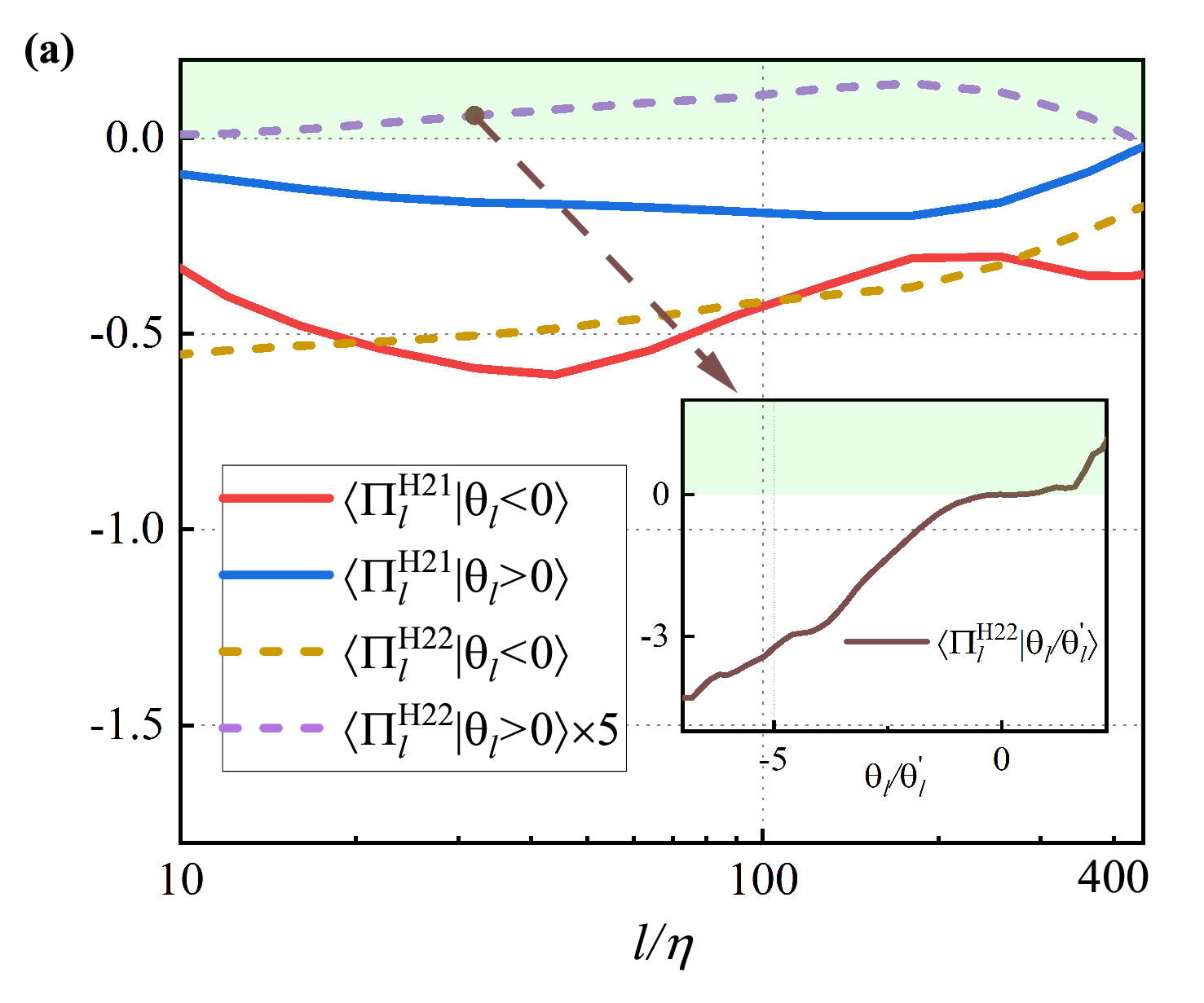}
		\end{minipage}%
	}
	
	\subfigure{
		\begin{minipage}[t]{0.9\linewidth}
			\centering
			\includegraphics[width=\textwidth]{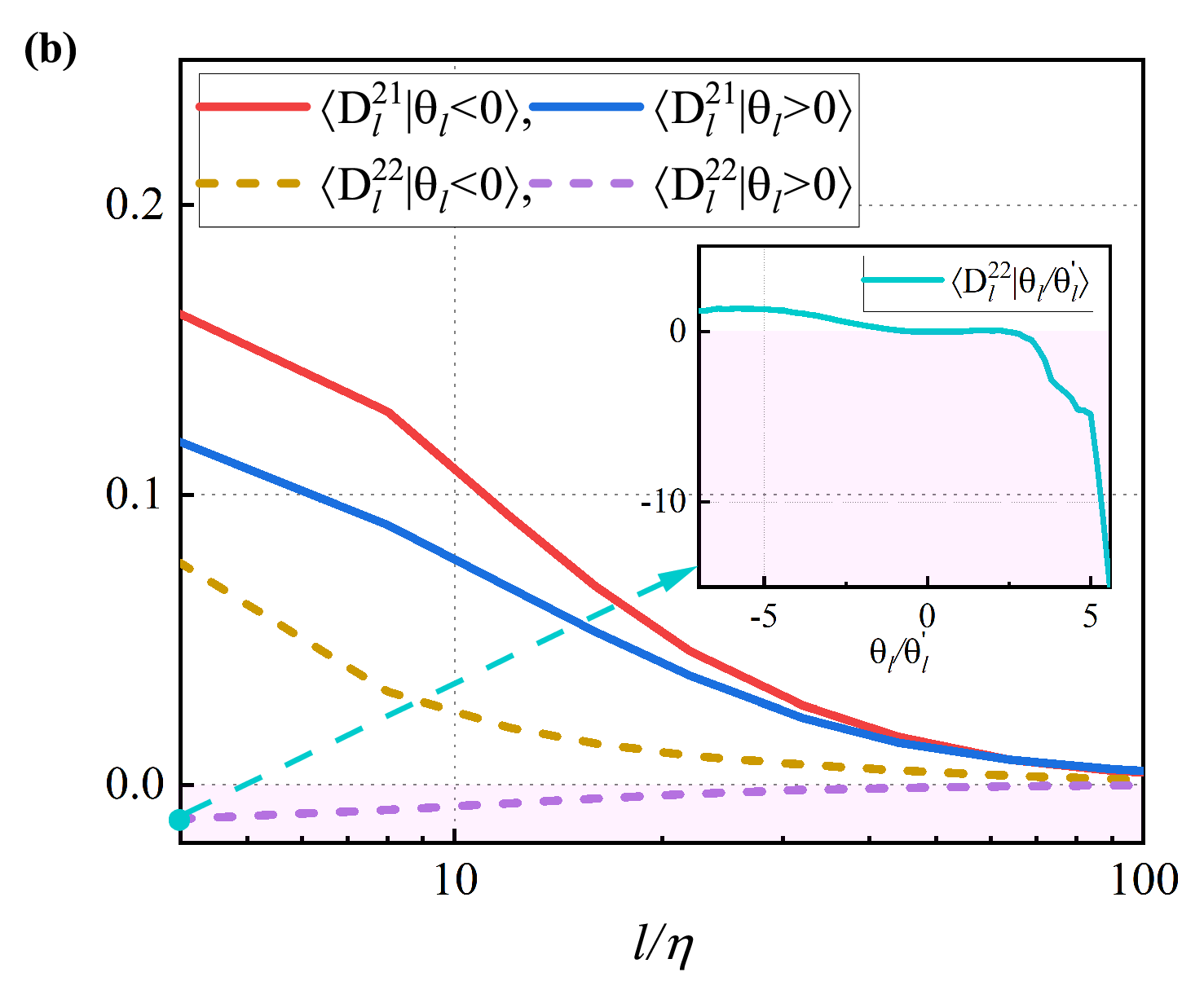}
		\end{minipage}%
	}
	\caption{(a) The ensemble averages of the first and second terms of the second-channel helicity flux conditioned in the compression $\theta_l<0$ and expansion regions $\theta_l>0$. The inset shows the ensemble average of the second term of the second-channel helicity flux conditioned in local velocity divergence with filter widths $l/\eta$=32. Here, $\theta_l=\partial \overline{u}_j/\partial x_j$. The light green region corresponds to the inverse cross-scale transfer phenomenon. (b) The ensemble averages of the first and second terms of the second-channel viscosity term conditioned in compression and expansion regions. The inset shows the ensemble average of the second term of the second-channel viscosity term conditioned in local velocity divergence with filter widths $l/\eta$=4. The light pink region corresponds to the source for the generation of helicity.}
	\label{fig:dual_channel}
\end{figure}

In figure~\ref{fig:dual_channel}(a), we show the ensemble averages of the first and second terms of the second-channel helicity flux conditioned in compression and expansion regions. Flows in the compression regions dominate the first and second terms of the second-channel helicity flux, which is consistent with the dominant forward helicity cascade process. However, the sign of the second term in the expansion regions is positive, and it corresponds to an inverse helicity cascade process. The expansion effects on the second term are further evaluated through ensemble averages conditioned in local velocity divergence, and the statistical results are exhibited in the inset in figure~\ref{fig:dual_channel}(a). Compression and expansion have opposite effects on the second term of the second-channel helicity flux. With an increase in expansion, the magnitude of the second term is larger, which indicates a stronger inverse helicity cascade process. Similarly, with an increase in compression, the magnitude of the second term also increases, and it corresponds to a stronger forward helicity cascade process. Hence, we can conclude that the second term of the second-channel helicity flux in expansion regions serves as an intrinsic mechanism for the inverse helicity cascade process, which might help to sustain large-scale helical structures. The statistical results of the first- and second-channel helicity fluxes conditioned in compression and expansion regions are consistent with the whole forward helicity cascade process, and we do not show them for the sake of simplicity.

The second term of the second-channel helicity flux in expansion regions provides an inverse cross-scale transfer mechanism at medium scales to sustain large-scale helical structures. However, at small scales, the viscosity terms are dominant. Hence, we also employ the above conditional averaging method to evaluate the compression and expansion effects on the statistical characteristics of the involved viscosity terms. In figure~\ref{fig:dual_channel}(b), we exhibit the ensemble averages of the first and second terms of the second-channel viscosity term conditioned in the compression and expansion regions. With an increase in the lengths of the scales, their magnitudes decrease rapidly. This indicates that the selected definition of large-scale helicity is appropriate because it satisfies the "inviscid criterion" proposed by~\citet{aluie2011compressible,zhao2018inviscid}. Similar to the conclusions for SGS terms, only the sign of the second term of the second-channel viscosity term in the expansion regions is the same as the total helicity. This means that the second term serves as a source for helicity generation at small scales in expansion regions. The inset in figure~\ref{fig:dual_channel}(b) shows the strength of the compression and expansion effects on the second term of the second-channel viscosity term. When the expansion is strong, the magnitude of the second term of the second-channel viscosity term is very large. This validates the expansion effects on the source role of the second term of the second-channel viscosity term.

\begin{figure}[htb]
\centerline{\includegraphics[scale=0.3]{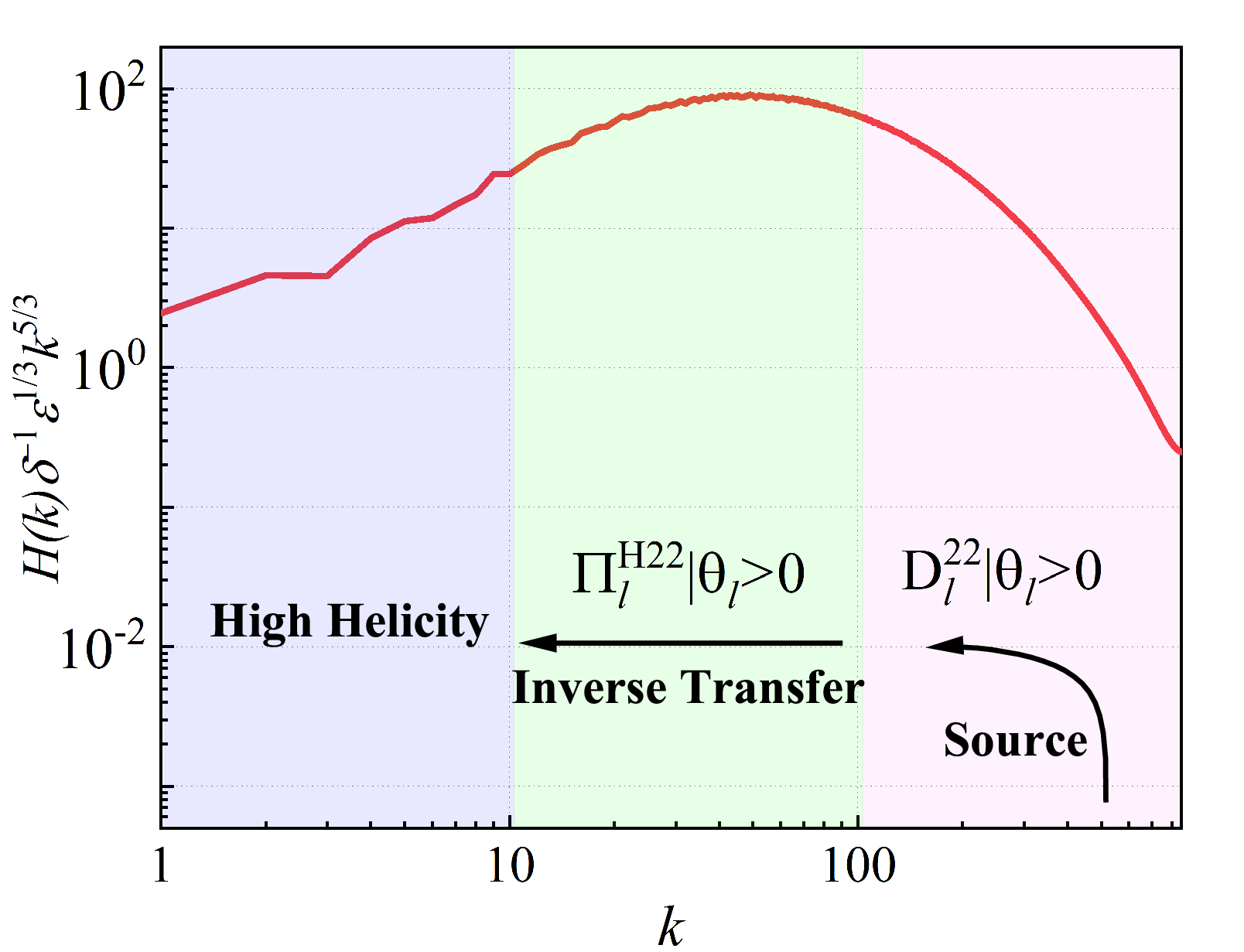}}
\caption{Normalized helicity spectrum.}
\label{fig:helicity_spectra_schematic_2048}
\end{figure}

In the above statistical analysis, we found that the second term of the second-channel helicity flux and the second term of the second-channel viscosity term in expansion regions serve as a self-induced mechanism for large-scale helicity. Both originate from the variable-density effects in the vorticity equation, which is neglected in incompressible turbulent flows. In fully developed helical turbulent flows, the helicity spectrum obeys a power-law solution as follows:
\begin{equation}
H(k) \sim C_H \delta \varepsilon^{-1 / 3} k^{-5 / 3}, 
\end{equation}
where $C_H$ is the Kolmogorov constant of the helicity spectrum \citep{chen2003intermittency,alexakis2018cascades,milanese2021dynamic}. In figure~\ref{fig:helicity_spectra_schematic_2048}, we show the normalized helicity spectrum in the present numerical simulations with a high Reynold number. Due to the limitation of the computational cost, the present Reynolds number is still not large to exhibit an apparent inertial subrange. Even the present Reynolds number is larger than our previous work ~\citep{yan2019effect,yan2020cross}. In addition, we schematically summarize the proposed induced mechanism in figure~\ref{fig:helicity_spectra_schematic_2048}. At small scales, the second term of the second-channel viscosity in expansion regions ($D_{\textit{l}}^{22}|\theta_{\textit{l}}>0$) serves as a source role for helicity generation. At medium scales, the second term of the second-channel helicity flux in expansion regions ($\Pi_{\textit{l}}^{H22}|\theta_{\textit{l}}>0$) serves as an inverse cross-scale transfer mechanism. It transfers helicity generated or deposited at small scales to large scales, and the large-scale helical structures can be sustained to some extent.

\section{Conclusions and discussions}

To summarize, we propose a self-induced mechanism for large-scale helical structures in compressible turbulent flows. The self-induction of large-scale helical structures originates from the intrinsic mechanism of its own system. The proposed self-induced mechanism provides an alternative theoretical foundation from the perspective of the intrinsic mechanism at inertial and small scales. In contrast to the dominant dissipation process, the self-induced mechanism is composed of the generation of helicity via the viscous process at small scales and the inverse cascade of helicity via scale interaction at inertial scales, which is associated with the expansion process of fluid elements. A high-order and high-resolution DNS with $2048^3$ verifies the accuracy of the present statistical analysis. 

To comprehensively investigate the statistical characteristics of helicity transfer in compressible turbulent flows, we select a suitable definition for large-scale helicity in compressible turbulent flows to represent the conservative character of helicity transfer, getting rid of pressure and viscosity terms. Similar to inviscid criteria for large-scale kinetic energy in compressible turbulent flows \citep{aluie2011compressible,aluie2012conservative,aluie2013scale,zhao2018inviscid}, large-scale helicity should involve the Favre-filtered velocity and vorticity, which is defined as $\widetilde{H}=\widetilde{\textbf{\textit{u}}}\cdot\widetilde{\bm \omega}$. Its governing equation~\ref{eq:favre filtered helicity equation1} indicates that only SGS transfer exists in the inertial subrange, without considering pressure and viscosity terms. Based on a simple algebraic operation, we can prove that the two point-to-point pressure terms originating from momentum and vorticity governing equations are the same in compressible turbulent flows, and the dual-channel characteristics of SGS and viscosity terms are highlighted. Through the conditional averaging method, the expansion effect on the statistical characteristics of the SGS and viscosity terms are investigated, and the proposed self-induced mechanism is based on the above statistical analysis.

In engineering flows, this self-induced mechanism provides an explanation of high skin-friction and heat flux induced by near-wall helical structures in highspeed aircraft. This self-induced mechanism opens the way to a more comprehensive physical understanding of long-period characteristic of large-scale helical structures, widely existing in the atmosphere, ocean and turbomachinery.

\section{Acknowledgements}
This work was supported by the National Natural Science Foundation of China (NSFC Grants No.12302281 and No. 12072349), and the National Key Research and Development Program of China (2019YFA0405300 and 2020YFA0711800). Z. Y. is also supported by the China Postdoctoral Science Foundation (No. 2022M710459). L. F. W. is also supported by NSFC Grant No. 11975053.

\appendix
\section{Governing equation of the large-scale helicity $\overline{H}=\overline{\textbf{\textit{u}}}\cdot\overline{\bm \omega}$ in compressible turbulent flows}\label{app:direct definition of large-scale helicity}

Making filter operations on the governing equations of velocity and vorticity, the following governing equations of filtered velocity and vorticity could be obtained,
\begin{equation}
\label{eq:filtered velocity equation}
\frac{\partial \overline{u}_i}{\partial t}+\overline{u}_j\frac{\partial \overline{u}_i}{\partial x_j}+\overline{u}_i \overline{\theta}=-\frac{\partial \overline{\tau}_{ij}}{\partial x_j}-\overline{u_i \theta}-\overline{\frac{1}{\rho}\frac{\partial p}{\partial x_i}}+\frac{1}{Re}\overline{\frac{1}{\rho}\frac{\partial \sigma_{ij}}{\partial x_j}},
\end{equation}
\begin{equation}
\begin{split}
\label{eq:filtered vorticity equation}
\frac{\partial \overline{\omega}_i}{\partial t}+\overline{u}_j\frac{\partial \overline{\omega}_i}{\partial x_j}+\overline{\omega}_i \overline{\theta}= & \overline{\omega}_j\frac{\partial \overline{u}_i}{\partial x_j}-\frac{\partial \overline{\gamma}_{ij}}{\partial x_j}-\overline{\epsilon_{ijk}\frac{\partial}{\partial x_j}\left( \frac{1}{\rho}\frac{\partial p}{\partial x_k} \right)} \\ & +\frac{1}{Re}\overline{\epsilon_{ijk}\frac{\partial}{\partial x_j}\left( \frac{1}{\rho}\frac{\partial \sigma_{kl}}{\partial x_l} \right)},
\end{split}
\end{equation}
where $\overline{\tau}_{ij}$ is the subgrid-scale stress defined as $\overline{\tau}_{ij}=\overline{u_i u_j}-\overline{u}_i\overline{u}_j$, and $\overline{\gamma}_{ij}$ is the subgrid-scale vortex stretching stress defined as  $\overline{\gamma}_{ij}=\overline{u_j \omega_i}-\overline{u_i \omega_j}-\left(\overline{u}_j\overline{\omega}_i-\overline{u}_i\overline{\omega}_j\right)$.

According to the definition of $\overline{H}$, we can get the following governing equation of large-scale helicity $\overline{H}$ through combining \ref{eq:filtered velocity equation} and \ref{eq:filtered vorticity equation} as 
\begin{equation}
\label{eq:filtered helicity equation1}
\frac{\partial \overline{H}}{\partial t}+\frac{\partial \overline{J}_j}{\partial x_j}=-\overline{H}_{} \overline{\theta}-\Pi^{\overline{H}1}-\Pi^{\overline{H}2}+\overline{u_i \theta}\overline{\omega}_i+\overline{\Phi}+\overline{D}.
\end{equation}
Here, the spatial transport term $\overline{J}_j$, and the first and second channels of helicity cascade $\Pi^{\overline{H}1}$ and $\Pi^{\overline{H}2}$ are defined as 
\begin{equation}
\overline{J}_j=\overline{u}_j \overline{H}-\left( 1/2 \right)\overline{u}_i \overline{u}_i \overline{\omega}_j + \overline{\omega}_i \overline{\tau}_{ij} + \overline{u}_i \overline{\gamma}_{ij},
\end{equation}
\begin{equation}
\Pi^{\overline{H}1}=\overline{\tau}_{ij}\frac{\partial \overline{\omega}_i}{\partial x_j}, \quad \Pi^{\overline{H}2}=\overline{\gamma}_{ij}\frac{\partial \overline{u}_i}{\partial x_j}.
\end{equation}
$\overline{\Phi}$ referring to pressure term and $\overline{D}$ referring to viscous term are defined as 
\begin{equation}
\overline{\Phi}=-\overline{\frac{1}{\rho}\frac{\partial p}{\partial x_i}} \overline{\omega}_i+ \overline{\frac{\epsilon_{ijk}}{\rho^2}\frac{\partial \rho}{\partial x_j} \frac{\partial p}{\partial x_j}} \overline{u}_i,
\end{equation}
\begin{equation}
\overline{D}=\frac{1}{Re}\overline{\frac{1}{\rho}\frac{\partial \sigma_{ij}}{\partial x_j}}\overline{\omega}_i+\frac{1}{Re}\overline{\epsilon_{ijk}\frac{\partial}{\partial x_j}\left( \frac{1}{\rho}\frac{\partial \sigma_{kl}}{\partial x_l} \right)} \overline{u}_i.
\end{equation}
The first term on the right hand side (r.h.s.) of equation~\ref{eq:filtered helicity equation1} only includes large-scale flow field, and it act as a resource term at all scales similar to vortex stretching term of enstrophy governing equation in three-dimensional turbulent flows. The presence of large-scale source term originating from velocity field breaks the conservative characteristic of helicity cascade, except pressure and viscosity terms in compressible turbulent flows. The other terms on the r.h.s. of equation~\ref{eq:filtered helicity equation1} involve large- and small-scale flow field. Especially, the two expressions of the inter-scale helicity are the same as those in incompressible turbulent flows.

\section{Tensor geometry of the first term of the second-channel helicity cascade}\label{app:first-term tensor geometry}

 For the convenience of tensor analysis, we introduce a density-weighted velocity $\textbf{\textit{v}}=\sqrt{\rho}\textbf{\textit{u}}$ and a filtered density-weighted velocity $\textbf{\textit{w}}=\sqrt{\overline{\rho}}\widetilde{\textbf{\textit{u}}}$. This weighted velocity form is suitable to investigate the spectral properties of physical variables in compressible flows. The SGS stress can be rewritten as$
\overline{\rho}\widetilde{\tau}_{ij}\equiv\overline{\sqrt{\rho}u_i \cdot \sqrt{\rho}u_j}-\sqrt{\overline{\rho}}\widetilde{\textit{u}}_i \cdot \sqrt{\overline{\rho}}\widetilde{\textit{u}}_j\equiv\overline{v_i v_j}-w_i w_j$.
Using above definitions, the tensor operation of the first term of the SGS stress involved in the first term can be written as 
\begin{equation}
\begin{split}
\epsilon_{ijk}\frac{\partial}{\partial x_j} &\left[\frac{\partial}{\partial x_m}\left( \overline{v_k v_m}\right)\right] \equiv \epsilon_{ijk}\frac{\partial}{\partial x_j}\left( \overline{v_m \frac{\partial v_k}{\partial x_m}}+ \overline{v_k \frac{\partial v_m}{\partial x_m} }\right) \\  & \equiv \overline{v_j\frac{\partial \omega_i}{\partial x_j}}- \overline{\omega_j\frac{\partial u_i}{\partial x_j}}+2\overline{\omega_i\theta} +\overline{\epsilon_{ijk}\frac{\partial \theta}{\partial x_j} v_k} \\ & \equiv \frac{\partial}{\partial x_j}\left( \overline{\omega_i v_j} -\overline{v_i \omega_j} \right)+\overline{\omega_i \theta}+ \overline{\epsilon_{ijk}\frac{\partial \theta}{\partial x_j} v_k},
\end{split}
\end{equation}
where $\bm\omega = \nabla \times \textbf{\textit{v}}$, and $\theta=\nabla \cdot \textbf{\textit{v}}$. Similarly, the tensor operation of the second term of the SGS stress involved in the first term can be written as
\begin{equation}
\begin{split}
\epsilon_{ijk}\frac{\partial}{\partial x_j} &\left[\frac{\partial}{\partial x_m}\left( w_k w_m\right)\right] \equiv \epsilon_{ijk}\frac{\partial}{\partial x_j}\left( w_m \frac{\partial w_k}{\partial x_m} + w_k \frac{\partial w_m}{\partial x_m} \right) \nonumber\\  & \equiv  w_j\frac{\partial \varpi_i}{\partial x_j}- \varpi_j\frac{\partial w_i}{\partial x_j}+2\varpi_i\vartheta +\epsilon_{ijk}\frac{\partial \vartheta}{\partial x_j} w_k \\  & \equiv \frac{\partial}{\partial x_j}\left( \varpi_i w_j -w_i \varpi_j \right)+\varpi_i \vartheta+ \epsilon_{ijk}\frac{\partial \vartheta}{\partial x_j} w_k,
\end{split}
\end{equation}
where $\bm\varpi = \nabla \times \textbf{\textit{w}}$, and $\vartheta=\nabla \cdot \textbf{\textit{w}}$.
Hence, the SGS term involved in the first term can be gotten
\begin{widetext}
\begin{equation}
\begin{split}
\epsilon_{ijk}\frac{\partial}{\partial x_j}\left[\frac{\partial}{\partial x_m}\left( \overline{v_k v_m} - w_k w_m \right)\right]  & = \frac{\partial}{\partial x_j}\left[ \left(\overline{\omega_i v_j} - \varpi_i w_j \right)-\left(\overline{v_i \omega_j}-w_i \varpi_j\right) \right] + \overline{\omega_i \theta}- \varpi_i \vartheta + \overline{\epsilon_{ijk}\frac{\partial \theta}{\partial x_j} v_k}- \epsilon_{ijk}\frac{\partial \vartheta}{\partial x_j} w_k
\end{split}
\end{equation}
\end{widetext}
We define $\Upsilon_{ij}=\left(\overline{\omega_i v_j} - \varpi_i w_j \right)-\left(\overline{v_i \omega_j}-w_i \varpi_j\right)$. It is an antisymmetric tensor, which is similar to the form in incompressible turbulent flows \citep{yan2020dual}.

\nocite{*}

\bibliography{Main}

\end{document}